\documentclass[fleqn,usenatbib]{mnras}

\usepackage{newtxtext,newtxmath}
\usepackage[T1]{fontenc}

\DeclareRobustCommand{\VAN}[3]{#2}
\let\VANthebibliography\thebibliography
\def\thebibliography{\DeclareRobustCommand{\VAN}[3]{##3}\VANthebibliography}

\usepackage{graphicx}	
\usepackage{amsmath}	
\usepackage{amssymb}	

\newcommand{\PAR}{Paranal Observatory}

\newcommand{\sigY}{$\sigma_{\scriptscriptstyle \text{Y}}$}
\newcommand{\sigYmath}{\sigma_{\scriptscriptstyle \text{Y}}}
\newcommand{\Cy}{$C_{\scriptscriptstyle \text{Y}}$}
\newcommand{\CyMath}{C_{\scriptscriptstyle \text{Y}}}
\newcommand{\Gmag}{$G_{\text{mag}}$}    
\usepackage{cprotect}   
\usepackage{cleveref}   
\usepackage{caption}    
\usepackage{subcaption} 

\usepackage{xcolor}
\usepackage{hyperref}
\newcommand{\orc}{\includegraphics[height=\fontcharht\font`A]{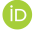}}
\newcommand{\orcid}[1]{\href{https://orcid.org/#1}{\orc}}

\title[Scintillation-limited  photometry with NGTS]{Scintillation-limited photometry with the 20-cm NGTS telescopes at Paranal Observatory}

\author[S. M. O'Brien et al.]{
Sean M. O'Brien\orcid{0000-0001-7367-1188},$^{1,2,3}$\thanks{E-mail: sobrien27@qub.ac.uk}
Daniel Bayliss\orcid{0000-0001-6023-1335},$^{1,2}$
James Osborn,$^{4}$
Edward M.~Bryant\orcid{0000-0001-7904-4441},$^{1,2}$
\newauthor
James McCormac,$^{1,2}$
Peter J. Wheatley\orcid{0000-0003-1452-2240},$^{1,2}$
Jack S.~Acton,$^{5}$
Douglas~R.~Alves\orcid{0000-0002-5619-2502},$^{6}$
\newauthor
David~R.~Anderson\orcid{0000-0001-7416-7522},$^{1,2}$
Matthew~R.~Burleigh,$^{5}$
Sarah~L.~Casewell,$^{5}$
Samuel Gill,$^{1,2}$
Michael~R.~Goad,$^{5}$
\newauthor
Beth~A.~Henderson,$^{5}$
James~A.~G.~Jackman,$^{7,1,2}$
Monika~Lendl,$^{8}$
Rosanna~H.~Tilbrook,$^{5}$
Stéphane Udry,$^{8}$
\newauthor
Jose~I.~Vines\orcid{0000-0002-1896-2377}$^{6}$
and
Richard G. West$^{1,2}$
\\
$^{1}$Department of Physics, University of Warwick, Gibbet Hill Road, Coventry, CV4 7AL, UK\\
$^{2}$Centre for Exoplanets and Habitability, University of Warwick, Gibbet Hill Road, Coventry CV4 7AL, UK\\
$^{3}$Astrophysics Research Centre, School of Mathematics and Physics, Queen's University Belfast, Belfast, BT7 1NN, UK\\
$^{4}$Centre for Advanced Instrumentation, Department of Physics, Durham University, South Road, Durham DH1 3LE, UK\\
$^{5}$School of Physics and Astronomy, University of Leicester, University Road, Leicester, LE1 7RH, UK\\
$^{6}$Departamento de Astronom\'ia, Universidad de Chile, Casilla 36-D, Santiago, Chile\\
$^{7}$School of Earth and Space Exploration, Arizona State University, Tempe, AZ 85287, USA\\
$^{8}$University of Geneva, Geneva Observatory, Chemin Pegasi 51, 1290 Versoix, Switzerland\\
}

\date{Accepted XXX. Received YYY; in original form ZZZ}

\pubyear{2021}

\begin{document}

\label{firstpage}
\pagerange{\pageref{firstpage}--\pageref{lastpage}}
\maketitle

\begin{abstract}
Ground-based photometry of bright stars is expected to be limited by atmospheric scintillation, although in practice observations are often limited by other sources of systematic noise. We analyse 122 nights of bright star ($G_{\text{mag}} \lesssim 11.5$) photometry using the 20-cm telescopes of the Next-Generation Transit Survey (NGTS) at the Paranal Observatory in Chile. We compare the noise properties to theoretical noise models and we demonstrate that NGTS photometry of bright stars is indeed limited by atmospheric scintillation. We determine a median scintillation coefficient at the \PAR\ of $\CyMath = 1.54$, which is in good agreement with previous results derived from turbulence profiling measurements at the observatory. We find that separate NGTS telescopes make consistent measurements of scintillation when simultaneously monitoring the same field. Using contemporaneous meteorological data, we find that higher wind speeds at the tropopause correlate with a decrease in long-exposure ($t=10$\,s) scintillation. Hence the winter months between June and August provide the best conditions for high precision photometry of bright stars at the \PAR. This work demonstrates that NGTS photometric data, collected for searching for exoplanets, contains within it a record of the scintillation conditions at Paranal.\\
\end{abstract}




\section{Introduction}\label{sec:intro}
High-precision time-series photometry is crucial to modern observational astronomy, in particular for astereoseismology \citep{Brown1994ARA&A..32...37B,Heasley1996PASP..108..385H} and transiting exoplanet research \citep{Winn2010exop.book...55W}.  The amplitudes of the signals that need to be detected extend down to hundreds of parts-per-million for Solar-type astereoseismic oscillations \citep{chaplin},  Earth-sized transiting exoplanets \citep{Winn2010exop.book...55W}, and Jupiter-sized secondary eclipses \citep{Knutson2007Natur.447..183K}.  Early transiting exoplanet surveys such as WASP \citep{pollacco}, HATNet \citep{hatnet}, and KELT \citep{kelt} used small ($\sim$5-10\,cm) ground-based telescopes to survey large areas of sky.  These surveys were typically limited to detecting signals on the order of 1\%.  \@ Larger telescopes ($\sim$20\,cm) were employed by the next generation of ground-based transit surveys such as NGTS  \citep{Wheatley2018MNRAS.475.4476W} and HATSouth \citep{hatsouth}.  Concurrently, missions such as CoRoT \citep{corot}, Kepler \citep{borucki}, and TESS \citep{Ricker2015JATIS...1a4003R} were able to reach new levels of photometric precision by operating in space.

The first generation of ground-based transit surveys suffered from correlated noise, which was the dominant source of noise in time-series photometry \citep[e.g.][]{Pont2006MNRAS.373..231P,Winn2007AJ....134.1707W,Cubillos2017AJ....153....3C}. The limiting factors are usually a combination of atmospheric conditions, telescope tracking and flatfield errors. Small telescopes usually suffer from imperfect tracking while larger telescopes have systematic noise due to a lack of suitable references stars. NGTS has overcome these issues by using a wide field-of-view ($2.8^{\circ} \times 2.8^{\circ}$), precise autoguiding system \citep[DONUTS][]{McCormac2013PASP..125..548M}; and high-quality back-illuminated CCDs \citep{Wheatley2018MNRAS.475.4476W}. In this paper we show that these improvements have lead to photometry that now reaches the scintillation limit (see Section~\ref{sec:ngts} for further details on NGTS).

When we observe from the ground, light from stars is distorted as it passes through turbulent regions of the Earth's atmosphere. Differences in air temperature in the atmosphere, due to circulation and Solar heating, cause differences in the density and therefore refractive indices of these regions. This in turn causes phase distortion of a plane light wave passing through the turbulent layers. As the light propagates through the atmosphere, the distortion increases and the wavefront interferes with itself resulting in both phase and intensity fluctuations. The spatial distortion of the point-spread function due to phase aberrations is the familiar concept of `seeing' which can be corrected through the use of adaptive optics \citep{Babcock1953PASP...65..229B}. However, when observing bright stars the dominant noise source can be atmospheric scintillation. Scintillation is the resulting variations in intensity of the light received by a telescope, due to the effects of the atmosphere \citep[e.g.][]{Dravins1997aPASP..109..173D}. These variations are typically of the order ~0.1--1.0\% \citep{Osborn2015MNRAS.452.1707O,Fohring2019MNRAS.489.5098F} which is similar to the depth of exoplanet transits or asteroseismology signals. Scintillation is more colloquially known as the `twinkling' we see when observing stars with the naked eye.

Efforts have been made to develop scintillation correction techniques \citep{Osborn2011MNRAS.411.1223O,Viotto2012SPIE.8447E..6XV,JOsborn2015MNRAS.446.1305O,Dhillon2016SPIE.9908E..0YD}.

\begin{figure}
    \centering
    \includegraphics[width=\columnwidth]{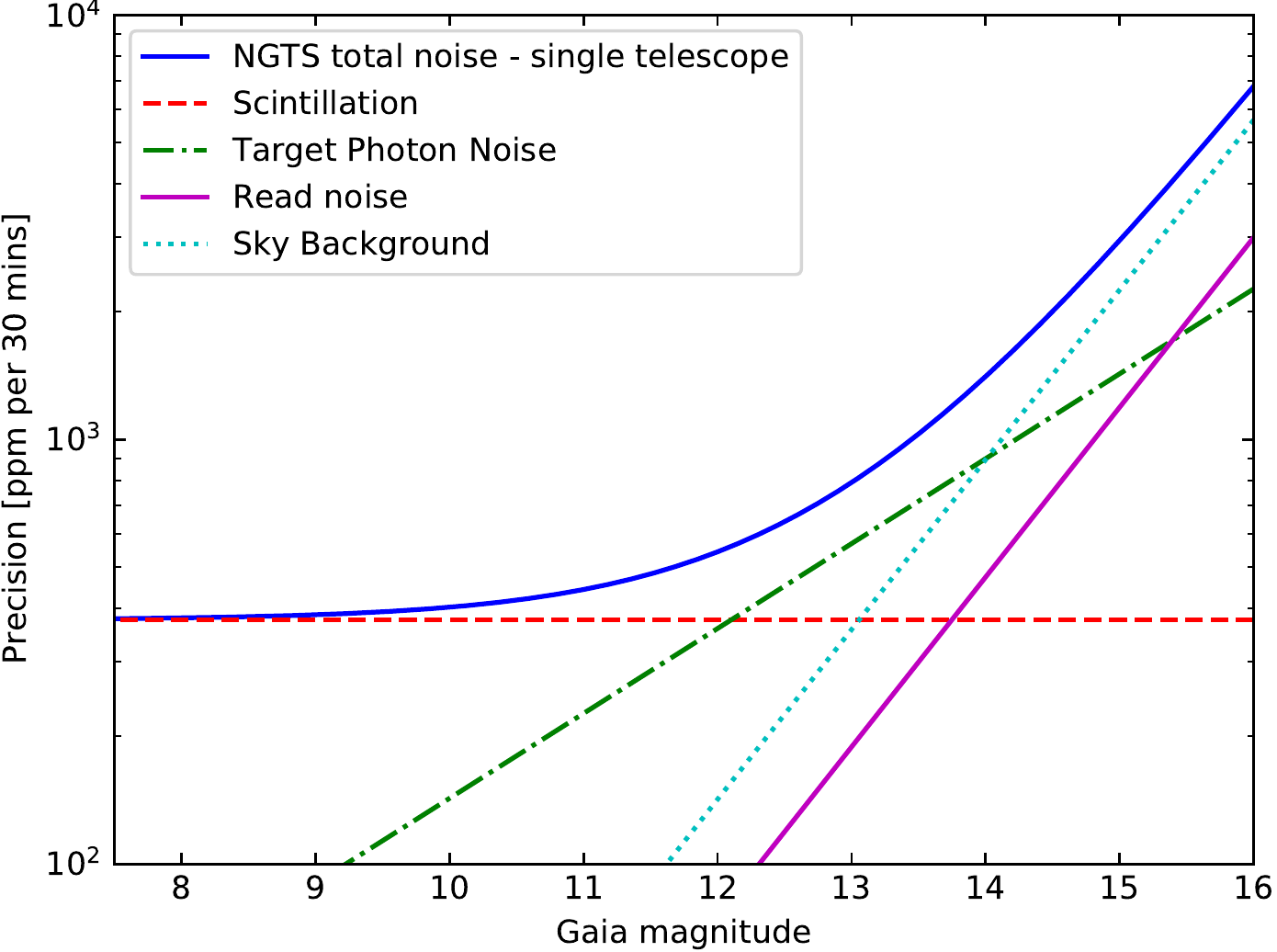}
    \caption{Theoretical noise model for a single NGTS telescope, including scintillation noise, target photon noise, sky background, and read noise.  The total noise (blue solid line) is the summation of these four components. Scintillation is expected to dominate the bright end, target photon noise dominates intermediate magnitudes and sky background dominates the faint end.}
    \label{fig:ngts_noise}
\end{figure}

In this paper, we analyse NGTS photometric data for a sample of around 22,000 observations of bright stars. Observations consist of the monitoring of a bright star in the field-of-view of an NGTS telescope for more than 2.4\,hours on a given night.  To define bright stars we use Gaia magnitudes \citep[\Gmag ;][]{Gaia2016A&A...595A...1G}, selecting stars with $G_{\text{mag}} \lesssim 11.5$.  This threshold magnitude was selected to match where the photometric noise is dominated by the scintillation effect for NGTS photometry \citep{Wheatley2018MNRAS.475.4476W}. At magnitudes fainter than this, NGTS photometry is limited by a combination of target photon noise, sky background and read noise, see Figure~\ref{fig:ngts_noise}.

The fractional amplitude of photometric noise due to scintillation (\sigY) can be estimated using the modified Young's approximation \citep{Young1967AJ.....72..747Y,Osborn2015MNRAS.452.1707O},

\begin{equation}\label{eq:modyoung}
   \sigYmath^2 = 10 \times 10^{-6} \CyMath^{2} D^{-4/3} t^{-1} (\sec z)^{3} \exp{(-2 h_{\text{obs}}/H)},
\end{equation}
where $D$ is the diameter of the telescope aperture (m), $t$ is exposure time used (s), $h_{\text{obs}}$ is the altitude of the observatory (2440\,m for Paranal), $H$ is the scale height of the atmospheric turbulence, taken to be 8000\,m. $z$ is the zenith distance, and therefore $\sec z$ approximates airmass. \Cy\ (m$^{2/3}$s$^{1/2}$) is the empirical coefficient.

This equation is taken from \citet{Osborn2015MNRAS.452.1707O}, where they introduced the empirical coefficient, \Cy, to improve the original approximation from \citet{Young1967AJ.....72..747Y}. This \Cy\ is typically found to be around $1.5$ for different sites, however this is a median value and the actual value can vary significantly across even a single night of observing. The median measured value of \Cy\ at the \PAR\ is $1.56^{+0.34}_{-0.29},$ where the limits are the upper and lower quartiles \citep{Osborn2015MNRAS.452.1707O}.

A more comprehensive measure of the scintillation index can be achieved using turbulence and wind velocity profilers. There exists two different scintillation regimes dependent on exposure time of the observations. In the long-exposure regime ($t \gtrsim 0.1$\,s), the intensity speckles which cause scintillation traverse the telescope pupil and scintillation reduces due to temporal averaging, with a dependency on high-altitude wind speed. In the short-exposure regime, this averaging does not occur \citep{Osborn2015MNRAS.452.1707O}. NGTS uses 10 second exposures and so operates in the long-exposure regime. The long-exposure scintillation index is given by Equation~4 in \citet{Osborn2015MNRAS.452.1707O}, it has the form
\begin{equation}\label{eq:LE_prop}
    \sigma_{I,\text{le}}^2 \propto \int_{0}^{\infty} \frac{h^2 C_n^2(h)}{V_\perp (h)} \text{d}h,
\end{equation}
where $C_n^2(h)$ is the profile of the refractive index structure constant, h is the altitude of the turbulent layer, $V_\perp (h)$ is the wind velocity profile. The $h^2$ term demonstrates that high altitude turbulence has a greater effect as scintillation is a propagation effect \citep{Osborn2015MNRAS.452.1707O}. The short-exposure scintillation index, Equations~3 and 6 in \citet{Osborn2015MNRAS.452.1707O}, do not have a wind velocity term. In this paper, we investigate whether there exists a seasonal correlation between wind speed and our long-exposure scintillation measurements, as expected based on the results presented in \citet{Kornilov2012A&A...546A..41K}.

Section~\ref{sec:data} describes the photometric data we collected from the NGTS facility. We also outline the data accessed from the MASS-DIMM and the European Centre for Medium-Range Weather Forecasts (ECMWF). In Section~\ref{sec:analysis} we analyse the NGTS scintillation data, comparing it to the MASS-DIMM and ECMWF datasets.  We describe the light curve analysis which calculates the root mean square (RMS) variability and the empirical scintillation coefficient \Cy.  We compare measurements from individual NGTS cameras as well as investigate seasonal variability in the scintillation data.  Finally in Section~\ref{sec:conclusions} we set out our conclusions from this study.


\section{Data}\label{sec:data}
\subsection{Photometry (NGTS)}\label{sec:ngts}
The Next Generation Transit Survey \citep[NGTS;][]{Wheatley2018MNRAS.475.4476W} is an array of twelve robotically-operated telescopes, each with 20 cm apertures and a field of view of $8\deg^2$. The telescopes are located at the \PAR\ in Chile, approximately 2\,km from the VLT and at an altitude of 2440\,m. NGTS uses a custom filter with a bandpass from 520 to 890\,nm. Each telescope is fitted with a $2048 \times 2048$ pixel CCD packaged into custom versions of the Andor iKon-L 936 camera. Read noise is around 14 electrons, the nominal exposure time is 10\,s with a 3\,s readout time, and thus a cadence of 13\,s \citep{Wheatley2018MNRAS.475.4476W}.

NGTS uses an updated version of the DONUTS autoguiding system to minimize star tracking issues and allow us to achieve high-precision photometry \citep{McCormac2013PASP..125..548M}. This stability means that stars are static on the pixels and there is no measurable red noise on short time scales.

The primary goal of NGTS was to survey large sections of the sky in search of exoplanet transits \citep[e.g.][]{Bayliss2018MNRAS.475.4467B}. Since 2018, NGTS has also been used for exoplanet follow-up observations of bright stars, particularly from the TESS mission \citep{Ricker2015JATIS...1a4003R}. The very wide field of view of the NGTS telescopes gives them a unique capability amongst ground-based telescopes of being able to reach 150\,ppm per 30 minutes for very bright stars \citep{Bryant2020MNRAS.494.5872B}. This has resulted in NGTS being used in the follow-up of bright stars, for example in refining the parameters for bright candidates from TESS \citep{Armstrong2020Natur.583...39A,Brahm2020AJ....160..235B} and monitoring transit timing variations for bright stars \citep{TTVBryant2021MNRAS.504L..45B}.

The data used in this paper were collected by the NGTS telescopes at the \PAR, Chile, between November 2018 and February 2021.

The data used are from specific bright star observations, which are performed in the same way as the standard survey mode, but may involve defocusing of the cameras and a reduced range of airmasses. This observation mode produces FITS images which are processed by a custom bright stars aperture photometry pipeline \citep{Bryant2020MNRAS.494.5872B}. This pipeline produces light curves for all of the bright stars in the field of view. The custom pipeline computes the photometry for a range of circular aperture radii. We use the photometry files with apertures of 6 pixels in radius, as this size of aperture captures the full extent of each star we wish to analyse. The bright star operation mode and pipeline are described by \citet{Bryant2020MNRAS.494.5872B}.

\subsection{Scintillation monitoring (MASS-DIMM)}\label{sec:mass}
The estimations of \Cy\ for the \PAR\ presented by \citet{Osborn2015MNRAS.452.1707O} are derived from measurements made using the MASS by \citet{Kornilov2012A&A...546A..41K}. The combined Multi-Aperture Scintillation Sensor (MASS)-Differential Image Motion Montior (DIMM) instruments are turbulence profilers at the \PAR\ which measure scintillation indices in order to monitor the atmosphere \citep{Kornilov2007MNRAS.382.1268K}. They use a short-exposure time of 1\,ms and have a 2\,cm aperture so the scintillation indices are approximated by Equation~6 of \citet{Osborn2015MNRAS.452.1707O}.
The MASS-DIMM target stars are very bright, $V \lesssim 2$, and are only observed at airmasses less than 1.5, therefore the target star switches during the night and we must split the NGTS data accordingly \citep{Kornilov2007MNRAS.382.1268K}. NGTS data products are numbered by an `actionID,' corresponding to the telescope ID, field observed and night of observation. We split `actions' into `sub-actions' if the MASS-DIMM instrument changes target star during the night because this allows analysis of correlations between target proximity and scintillation measurements.

\subsection{Wind speed data (ECMWF)}\label{sec:ecmwf}
The European Centre for Medium-Range Weather Forecasts (ECMWF) provides a plethora of data on the Earth's atmosphere and its weather systems. We use the `ERA5 monthly averaged data on pressure levels from 1979 to present' dataset which is a reanalysis of global weather that combines weather models with real observations from across the globe \citep{ECMWFdata}. We use this dataset at the latitude and longitude of the \PAR, however we note that the wind speed analysis could be carried out for any location, and therefore any observatory, on Earth. This work uses information on the U (eastward) and V (northward) wind vector components at a constant pressure level of 250\,hPa, which corresponds to where the subtropical jet stream lies at an altitude of approximately 12000\,m \citep{US1976ussa.rept......,ArcherJetstream}. We tested pressure levels of 200\,hPa and 225\,hPa but found no significant difference in the subsequent analysis. The jet stream, along with solar heating, injects energy into the atmosphere at the tropopause which leads to turbulence \citep{Kolmogorov1941DoSSR..30..301K,Roddier1981PrOpt..19..281R}. We can map the position of the jet stream above South America and this allows us to calculate a monthly, quantitative measure of the mean wind speed above Paranal, which is important when considering the changes in both short-exposure and long-exposure scintillation indices \citep{Kornilov2012A&A...546A..41K,Osborn2015MNRAS.452.1707O}.

\section{Analysis}\label{sec:analysis}
\subsection{Light curve analysis}\label{sec:pipeline}
We process the raw NGTS light curves (Section~\ref{sec:ngts}) by first filtering out poor nights of observation (i.e. clouds passing across the line-of-sight) and stars with light curves that do not follow similar trends to other stars in the same photometry file, for example variable stars. Additionally, stars with a measured flux below 30,000 ADU counts per exposure ($G_{\text{mag}} \approx 11.5$) are removed as noise terms other than scintillation begin to dominate \citep{Wheatley2018MNRAS.475.4476W}. Stars with flux values greater than 900,000 counts per exposure ($G_{\text{mag}} \approx 7.5$) are removed due to the likelihood that they will saturate the NGTS CCD pixels.
The light curves are normalized and then detrended by constructing a master reference star which is the sum of the fluxes for all stars in the frame minus the current target star. The `current target star' cycles through each star in the photometry file which has survived the filtering. Stars with a particularly high RMS value across the full light curve, in comparison with the other stars in its frame, are removed as a final automated check for variable stars.

\begin{figure}
    \centering{\phantomsubcaption\label{fig:rawlc}\phantomsubcaption\label{fig:normlc}\phantomsubcaption\label{fig:rmsplot}}
    \includegraphics[width=\columnwidth]{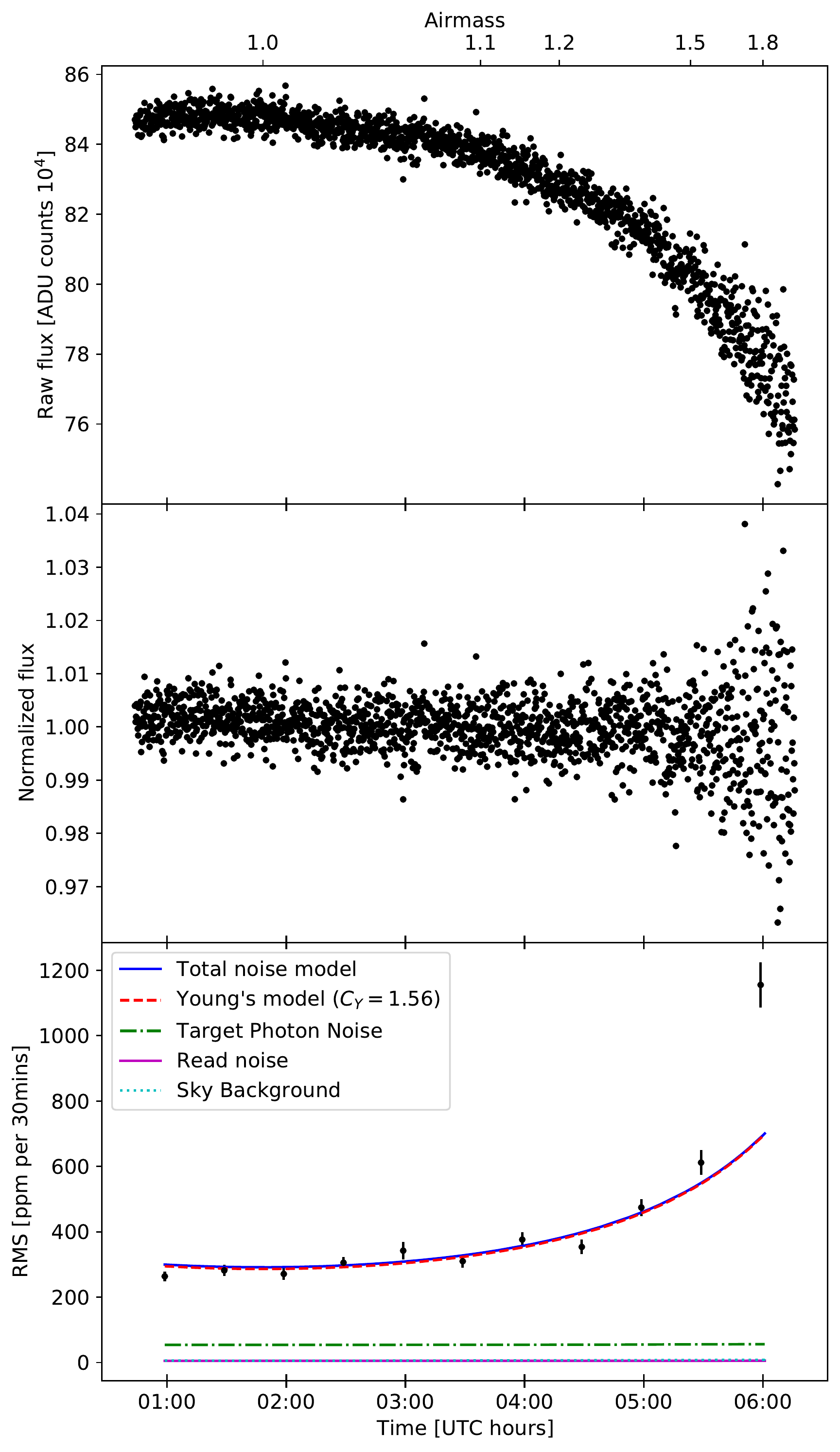}
    \caption{\textit{Top panel}, Figure~\ref{fig:rawlc}: Raw NGTS light curve for a star observed on 2020 January 28. This star has a Gaia magnitude of $G_{\text{mag}}=8.04$ therefore the photometric noise is expected to be dominated by scintillation. Airmass is displayed on the top axis. Airmass started at around 1.03, decreased to 1.0 at approximately 2\,am UTC, then increased to 1.98 at the end of the night. \textit{Middle panel}, Figure~\ref{fig:normlc}: Normalized light curve for the same star as above. Light curve is detrended and normalized via relative photometry. \textit{Bottom panel}, Figure~\ref{fig:rmsplot}: Photometric precision against time for the light curve shown above. The solid blue line is the total noise model (Eq.~\ref{eq:noise_model}), the dashed red line is the modified Young's equation (Eq.~\ref{eq:modyoung}) with the site median $\CyMath = 1.56$. The dark green dash-dot line is the target photon noise, the dotted cyan line is the sky background, the solid magenta line is the read noise. We note that the sky background and read noise almost coincide at approximately 50\,ppm per 30 mins. We also note that the scintillation component sits just below the total noise model because the model for this star is scintillation-dominated.}
    \label{fig:pipeline}
\end{figure}

In this section we display light curves and an RMS curve (\Cref{fig:rawlc,fig:normlc,fig:rmsplot}) taken from observations of a bright star (HD36109; $G_{\text{mag}} = 8.04$) on 2020 January 28. These data were taken during NGTS follow-up observations of TOI-431 \citep{Ares2021arXiv210802310O}.

The airmass trend is the dominant arching feature of the raw light curves, see Figure~\ref{fig:rawlc}. Higher airmass means we are looking through a larger column of air which leads to greater attenuation of light, as shown by the decrease in flux over the course of the night of observation. Furthermore, the greater spread of flux data points at higher airmasses, which is more apparent in the normalized light curve shown in Figure~\ref{fig:normlc}, is due to stronger scintillation. This increase in scintillation is accounted for by the $(\sec z)^3$ term in Equation~\ref{eq:modyoung}. The higher airmass, $\sec z$, means the distance to the turbulent layer is longer and so scintillation is increased as it is a propagation effect.

The next step is to calculate the RMS over 30 minute intervals across the normalized light curve. We show a plot of these values for the example light curve in Figure~\ref{fig:rmsplot}. The values and errorbars are calculated using bootstrap resampling \citep{AstropyI2013A&A...558A..33A,AstropyII2018AJ....156..123A}. We take 10,000 samples with replacement of the normalized flux values for each 30 minute interval and compute the standard deviation on each of these samples. The mean of the 10,000 standard deviations gives the RMS value for each 30 minute interval and the standard deviation of these standard deviations gives the errorbars.

Following \citet{McCormac2017PASP..129b5002M}, we calculate the total noise model as a combination of target photon noise, sky background, read noise and scintillation. Dark current is negligible as the NGTS cameras are sufficiently cooled to $-70^{\circ}$C \citep{Wheatley2018MNRAS.475.4476W}. The noise terms are computed in terms of ADU counts, therefore we must account for camera gain which ranges from 1.87 to 3.04 for the different NGTS cameras. The noise model used gives the total amount of noise, $N_T,$ in ADU counts as,
\begin{equation}\label{eq:noise_model}
    N_T^2 = N_{\text{target}}^2 + N_{\text{sky}}^2 + (n_{\text{pix}}\times N_{\text{read}}^2) + (\sigYmath \times f)^2.
\end{equation}
The target photon noise and sky background are characterised by photon-counting Poisson statistics. $n_{\text{pix}}$ is the number of pixels in a photometric aperture. $N_{\text{read}}$ is the read noise per pixel which is approximately 7 ADU counts, although it is camera-dependent. The final term is the scintillation term, using Equation~\ref{eq:modyoung} with the median value of $\CyMath=1.56$ for the \PAR. Since \sigY\ is a fractional uncertainty we multiply by source count rate, $f$, to express it in ADU counts. Each of these noise components, and the total noise model, are plotted as separate lines in Figure~\ref{fig:rmsplot}.

\subsection{Measuring the empirical scintillation coefficient}\label{sec:measurecy}
\begin{figure}
    \includegraphics[width=\columnwidth]{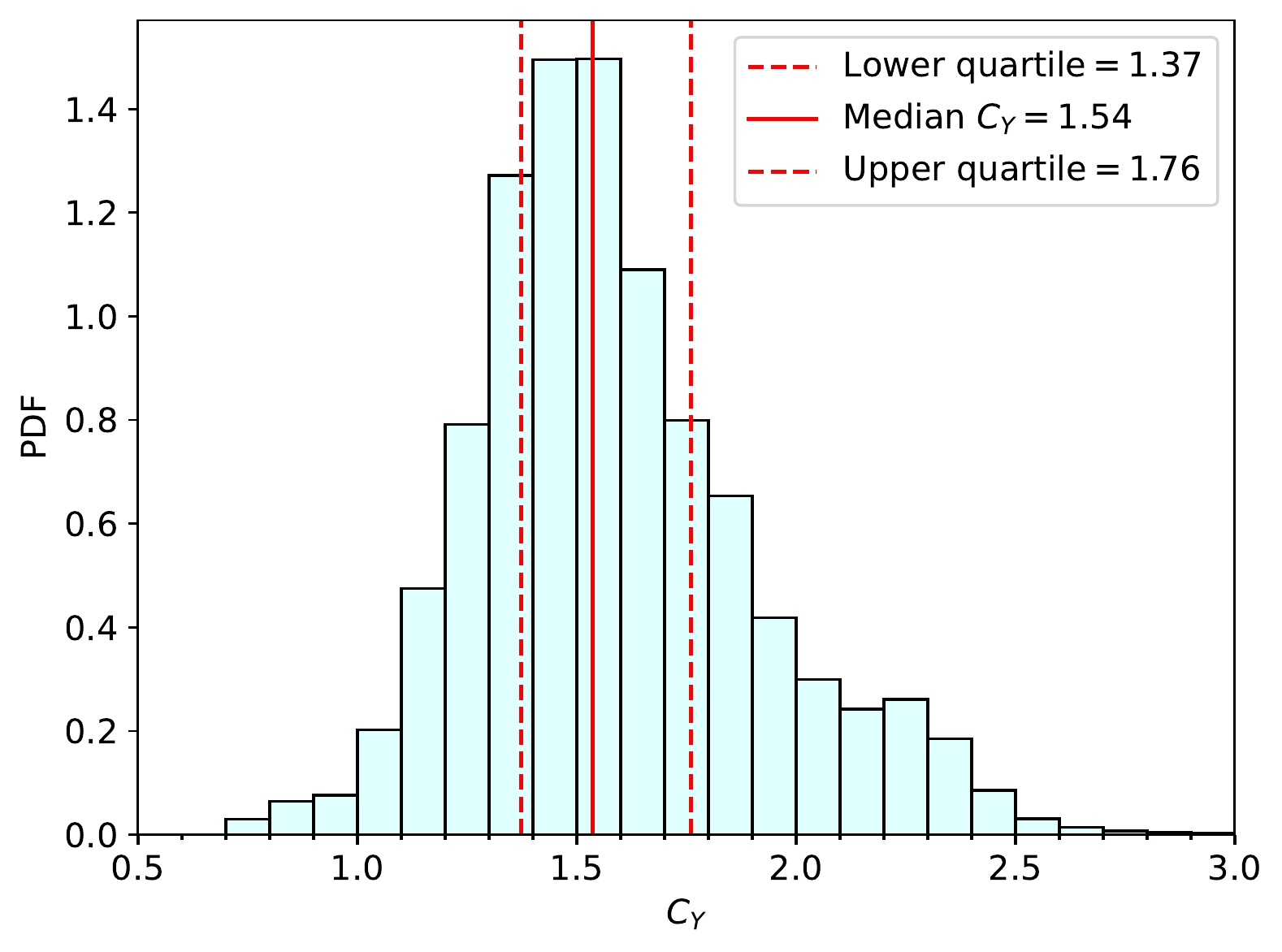}
    \caption{Normalized histogram of scintillation coefficients (\Cy) measured with the NGTS telescopes at the Paranal Observatory for all 21643 stars from 441 sub-actions across 122 nights. Values are weighted by $1/\sigma^2$, where $\sigma^2$ is the variance of the fit, and weights are normalized such that the area under the histogram equals 1. The solid red vertical line indicates the weighted median of the distribution ($\CyMath = 1.54$) and the dashed lines indicate the lower and upper quartiles of 1.37 and 1.76.}
    \label{fig:cy_hist}
\end{figure}

We fit our total noise model (Equation~\ref{eq:noise_model}) to the RMS curve for each star with the empirical scintillation coefficient, \Cy, as our only free parameter. We take the variance of the fit as our confidence level in the \Cy\ value. We plot a weighted histogram of all the \Cy\ values measured, using the reciprocal of the variance of the fit as the weightings. This is shown in Figure~\ref{fig:cy_hist}. We calculate a median of $\CyMath = 1.54$ with lower and upper quartiles of 1.37 and 1.76, respectively. These values are in close agreement with the median \Cy\ value of 1.56 and quartiles of 1.27 and 1.90 for the \PAR\ from \citet{Osborn2015MNRAS.452.1707O}. This agreement between independent methods suggests that NGTS photometry of bright stars is indeed limited by scintillation. The presence of other dominant systematic noise sources would mean that we would not be able to reach this limit.

\subsection{Dependence on camera}\label{sec:camvar}
\begin{figure*}
    \includegraphics[width=\textwidth]{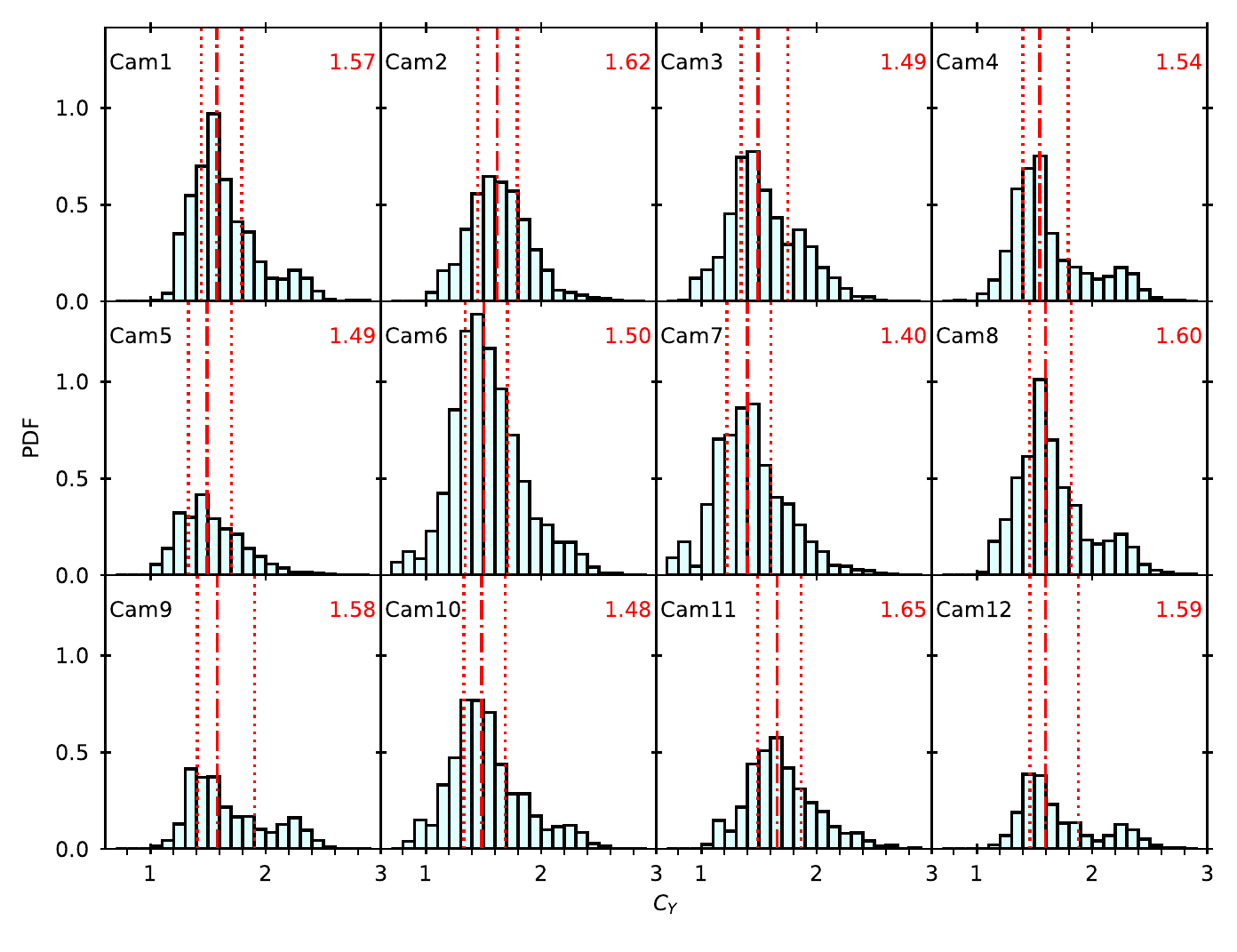}
    \caption{Normalized histograms displaying the \Cy\ values for all stars, split by camera. Each value used is weighted as in Section~\ref{sec:measurecy} and the weighted median is indicated by the red dash-dot lines, with the value displayed on each plot in red. The upper and lower quartiles for each camera are shown by vertical dotted lines. The histograms are normalized such that the relative size of histograms corresponds directly with the number of measurements, i.e. Cam6 took the most measurements for our sample.}
    \label{fig:camvar_1}
\end{figure*}

The NGTS facility provides us with the opportunity to measure any camera-dependent systematic noise by using our measurements of scintillation. Figure~\ref{fig:camvar_1} shows histograms of the scintillation measurements made by each camera. This figure shows that generally the 12 NGTS cameras have a similar distribution of \Cy\ values and there is no significant difference in the measurements made by the individual instruments at the facility.

We note there is a minor peak at $\CyMath=2.3$ for some telescopes. This is caused by a relatively large amount of data being collected over a few nights with high scintillation during observation periods that utilised only a subset of the telescopes.

We now combine \Cy\ values for all stars on each night for each MASS-DIMM target, but we treat each NGTS telescope measurement as an independent value. This gives one \Cy\ value per sub-action (Section~\ref{sec:ngts}). These \Cy\ values are computed as a weighted mean, using the weights as described in Section~\ref{sec:measurecy} and we calculate associated weighted standard deviations.

\begin{figure}
    \includegraphics[width=\columnwidth]{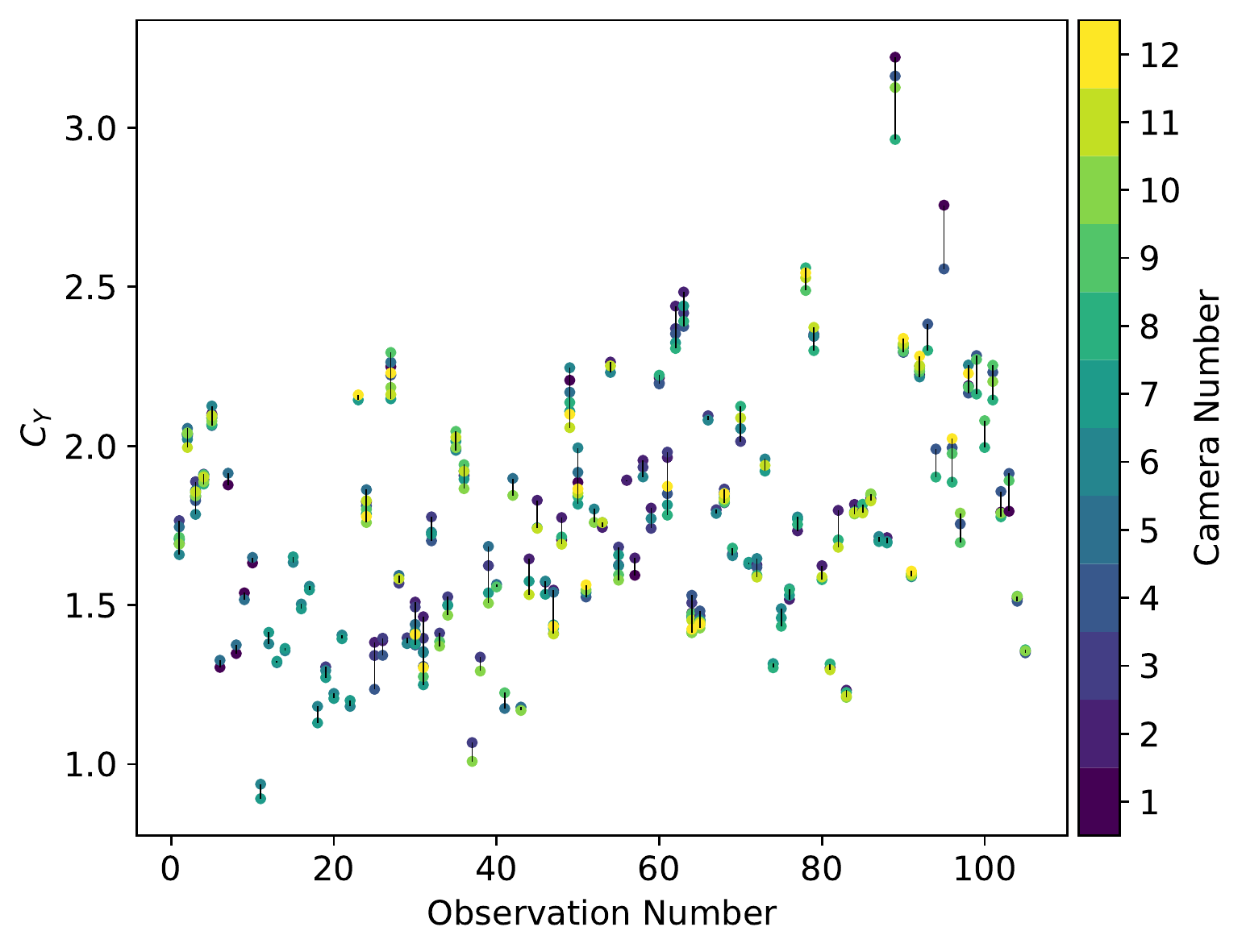}
    \caption{Scatter plot showing the value of \Cy\ for each sub-action, organised by night of observation. Only observation periods with multiple telescopes observing the same field are shown. Points are colour-coded by camera number using a discrete colourmap. Black vertical lines show the range of \Cy\ for each observation period.}
    \label{fig:camvar_2}
\end{figure}

Figure~\ref{fig:camvar_2} shows the inter-night variation between scintillation measurements made by the different NGTS cameras, also it shows the variation between nights of observation. This figure demonstrates that the difference between the values of \Cy\ measured by different telescopes is much less than the variation between different nights due to changes in the atmospheric turbulence. It also demonstrates that NGTS has overcome common issues such as lack of suitable reference stars, imperfect tracking or focusing errors to achieve scintillation-limited photometry.

\subsection{NGTS-MASS Correlation}\label{sec:ngts_mass}
\begin{figure}
    \includegraphics[width=\columnwidth]{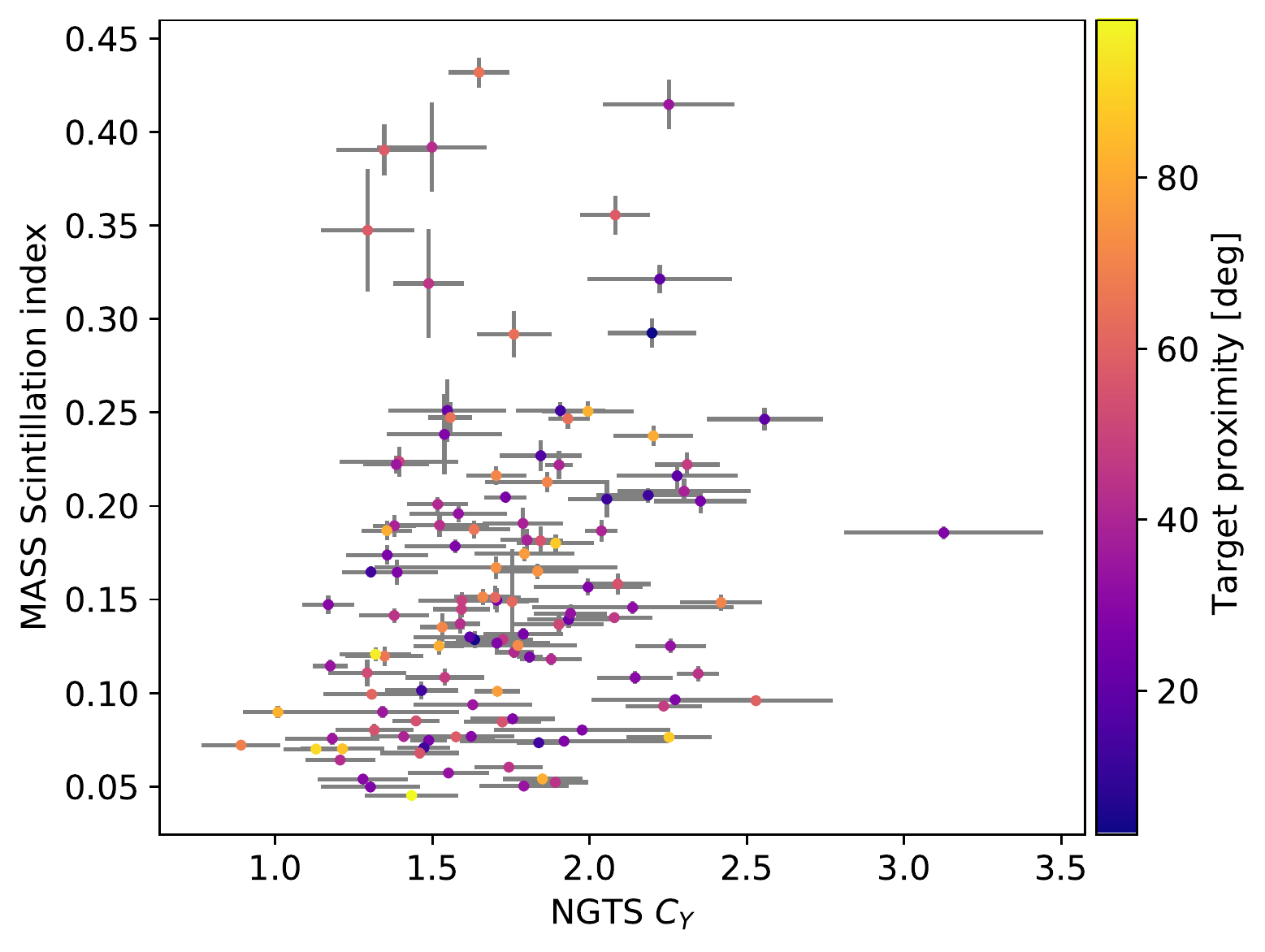}
    \caption{Scintillation coefficients (\Cy) measured by NGTS plotted against the scintillation indices measured by the MASS instrument. We plot all 122 nights for which NGTS observed bright stars, taking the median \Cy\ value for nights with multiple NGTS telescopes on the same target field. Colour-coded by separation between NGTS field center and MASS-DIMM target star.}
    \label{fig:ngtsvmass}
\end{figure}

The different scintillation exposure time regimes means that we do not anticipate correlation between the NGTS (long-exposure) and MASS-DIMM (short-exposure) measurements. To test this, we use the mean \Cy\ values for each NGTS sub-action computed in Section~\ref{sec:camvar}, taking the median value on nights with multiple NGTS observations. We compute the mean MASS scintillation measurement for each sub-action. The error on this mean MASS scintillation index is calculated as the standard error of the mean. We compare the correlation between the NGTS and MASS measurements by plotting the two datasets against each other, shown in Figure~\ref{fig:ngtsvmass}. The sky separation between the NGTS field center and MASS-DIMM target star, in degrees, is used to colour-code the plot.

As expected, we do not see strong correlation between the two measurements of scintillation since the MASS operates in the short-exposure regime while NGTS uses a long exposure time. We find a correlation coefficient of 0.22, indicating weak positive correlation. We do not see a stronger correlation for targets that are closer in proximity.

\subsection{Seasonal scintillation variability}\label{sec:longvary}
The ECMWF \citep{ECMWFdata} provides information on the global wind velocity field, therefore we can use these data to test the relation between scintillation and high-altitude wind speeds. In Section~\ref{sec:intro} we explained how higher altitude turbulence has a greater effect on the scintillation measured by astronomical instruments. At Paranal, the southern hemisphere subtropical jet stream is of particular interest since it is found close to $30^{\circ}$S at 10-16\,km above the surface, therefore we use the 250\,hPa pressure data as this corresponds to an altitude of around $h=12$\,km \citep{US1976ussa.rept......,ArcherJetstream}. The ERA5 dataset has horizontal resolution of $0.25^{\circ} \times 0.25^{\circ}$ so we take the $(U,V)$ wind components in the $1^{\circ} \times 1^{\circ}$ square centered on Paranal, $(\phi, \lambda) = (24.6^{\circ}\text{S},70.4^{\circ}\text{W}).$ We calculate the wind speed at each point in the data grid as $\sqrt{U^2+V^2}$ and compute the mean value of these wind speeds above the site for each month since January 2016.

\begin{figure}
    \includegraphics[width=\columnwidth]{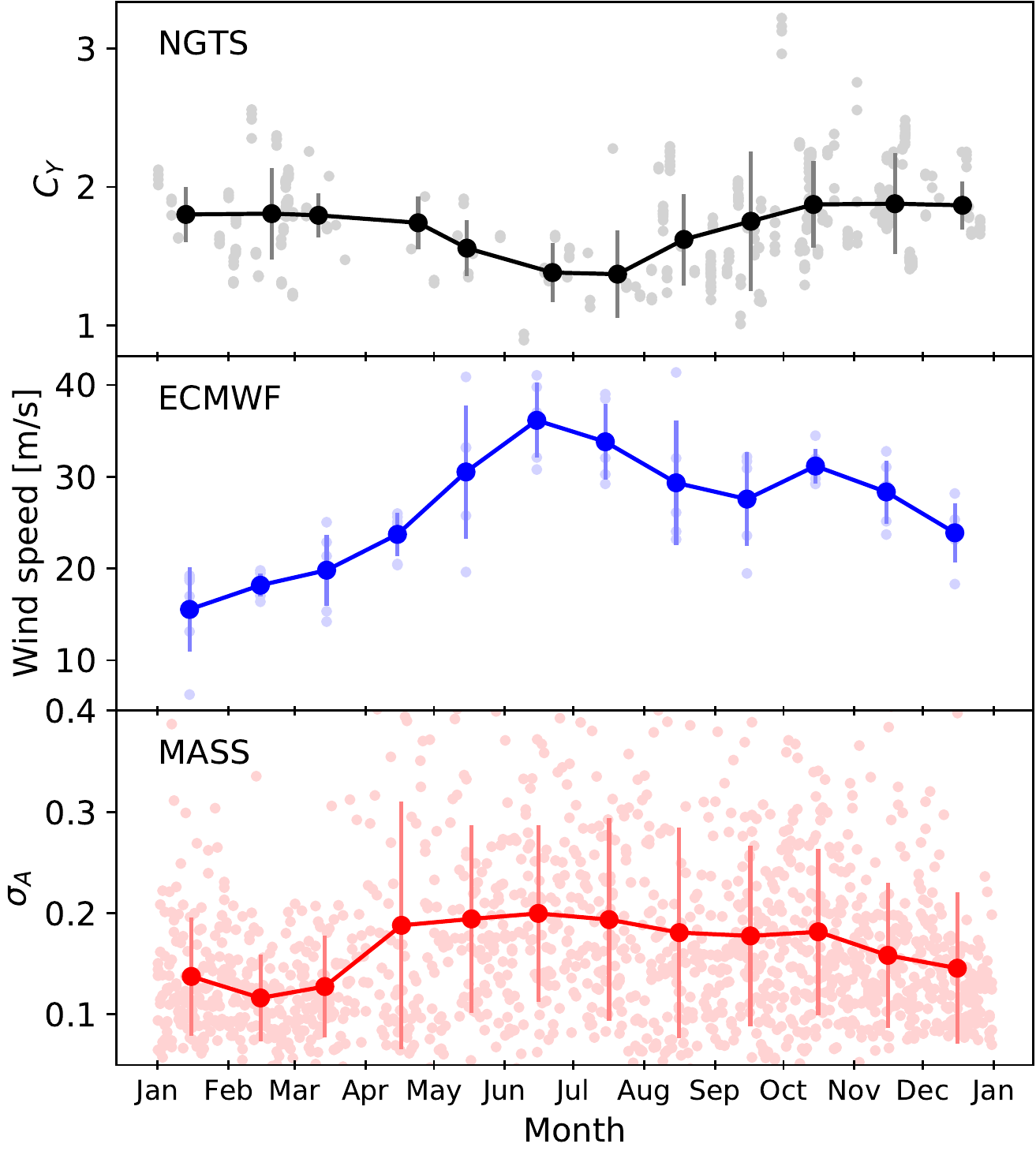}
    \caption{\textit{Top panel}: NGTS measurements of \Cy\ folded onto a one-year period. Light grey points are the measurements for each sub-action and black circles show the mean for each month, with standard deviation shown as grey errorbars. \textit{Middle panel}: ECMWF data for mean wind speed at 250\,hPa above Paranal. Faint blue points are the values for each month since 2016, blue circles are the mean values for each month across the full dataset with standard deviation shown as light blue errorbars. \textit{Bottom panel}: MASS measurements of the scintillation index. Faint red points are the mean measurements of $\sigma_A$ for each night (04/2016-03/2021), red circles show the monthly mean with standard deviation shown as light red errorbars.}
    \label{fig:mass_ngts_ecmwf_1y}
\end{figure}

Figure~\ref{fig:mass_ngts_ecmwf_1y} shows the annual variation in the NGTS \Cy\ measurements, the mean wind speed above Paranal, as computed from ECMWF data, and the MASS scintillation indices. The \Cy\ values for each sub-action as described in Section~\ref{sec:camvar} are displayed as light grey points. For each month we then compute the mean value and standard deviation. ECMWF data are acquired as monthly-averaged data from January 2016 through until April 2021. The mean of the full 4-5 years for each month is plotted, with corresponding standard deviation. We do note that the MASS data coverage is not uniform over these time periods due to operational interruptions.

We can see that the NGTS measurements of \Cy, which acts as a proxy for the long-exposure scintillation index from Equation~\ref{eq:LE_prop}, shows a minima in the months of June to August. The ECMWF wind speed measurements peak in these months. This indicates that long-exposure scintillation does indeed decrease when the wind speeds in the upper atmosphere are higher. This is because the higher wind speed averages out the turbulence cells passing across the line-of-sight of the telescope during the long exposure time. This causes an averaging out of the variation of light intensity and therefore reduces scintillation. This is governed in Equation~\ref{eq:LE_prop} as the $\frac{1}{V_{\perp}}$ term. We calculate a correlation coefficient of $-0.65,$ indicating moderately strong negative correlation between long-exposure scintillation and the mean wind speed.

It is evident from the bottom panel of Figure~\ref{fig:mass_ngts_ecmwf_1y} that the short-exposure scintillation index increases in the months of April to July. Additionally, the standard deviation, which indicates the variation in scintillation during the month, is much greater in these months. This suggests that scintillation conditions are much more unstable at the \PAR\ during these months. The peak in scintillation correlates with the peak in the high-altitude wind speeds measured by the ECMWF data. The higher wind speeds cause greater amounts of turbulence, therefore increasing $C_n^2(h)$. Meanwhile the short exposure time of the MASS measurements means that the averaging of the variation that occurs in the long-exposure regime is not a factor. The instability in scintillation conditions is likely due to higher wind speeds being more likely but less sustainable during these months \citep{ArcherJetstream}. We calculate a correlation coefficient of 0.86, indicating strong positive correlation between the MASS scintillation measurements and the mean wind speed.

\citet{Kornilov2012A&A...546A..41K} determined that long-exposure scintillation is minimal from May to September. Our conclusion is consistent with this. We also show the anti-correlation with high-altitude wind speed which is predicted by \citet{Kornilov2012A&A...546A..41K}. Similarly, short-exposure scintillation is maximal in the period from July to September. We find a similar result however the peak is closer to being from May to July. We also show the increase in short-exposure scintillation does correlate with wind speed, as predicted by \citet{Kornilov2012A&A...546A..41K}.

Figure~\ref{fig:mass_ngts_ecmwf_1y} also supports their conclusions that the seasonal variability does not exceed the quartiles (or in our case standard deviations) which indicates there is a reasonable probability of both good and poor conditions for bright star photometry year-round \citep{Kornilov2012A&A...546A..41K}.



\section{Conclusions}\label{sec:conclusions}
We analyse the photometric precision of two years of NGTS bright star observations from the \PAR.  We find that the dominant source of photometric noise for these stars is from atmospheric scintillation, and it is well described by the modified Young's equation for low airmass (airmass $< 1.5$). We find the median value for the empirical scintillation coefficient at the \PAR\ to be $\CyMath = 1.54$ with lower and upper quartiles of 1.37 and 1.76, respectively. This is in good agreement with the value of $\CyMath = 1.56$ derived by \citet{Osborn2015MNRAS.452.1707O} from MASS measurements taken by \citet{Kornilov2012A&A...546A..41K}.

All 12 NGTS telescopes have a similar distribution of measured scintillation coefficients, indicating that each individual telescope/camera reaches the scintillation limit when observing bright stars (see Figure~\ref{fig:camvar_1}). Additionally, we find that the 12 NGTS telescopes give consistent results for the measured scintillation when simultaneously observing the same field (Figure~\ref{fig:camvar_2}). This indicates that the telescopes and cameras are operating without any major source of camera-dependent systematic noise in this magnitude regime. This confirms that NGTS has overcome common issues in telescope design and operation to reach the scintillation limit for bright stars and provide high-precision photometry.

As expected, we see no strong correlation between the NGTS and MASS measurements of scintillation (Figure~\ref{fig:ngtsvmass}), owing to the fact that the instruments operate in different exposure time regimes (long-exposure and short-exposure regimes respectively).

Our work provides observational evidence for the seasonal variation cycle in both the long-exposure and short-exposure scintillation regimes. In the long-exposure regime, we find that scintillation is minimal from June to August and we provide evidence that this seasonal variation correlates with the peak in high-altitude wind speed above the \PAR. This is likely due to the turbulence cells being averaged out as they pass across the telescopes line-of-sight during the 10\,s exposure time of the NGTS cameras.
In the short-exposure regime, we find that scintillation is maximal from April to July. This peak matches with the peak in high-altitude wind speed which increases the turbulence strength, while the short-exposure time of the MASS instrument means there is no averaging out of the turbulence cells (see Figure~\ref{fig:mass_ngts_ecmwf_1y}).
We will continue to monitor the monthly variation in both scintillation regimes and study the correlation with wind speed. The analysis of wind speed variation and its correlation with long-exposure scintillation could be extended to other observing sites.

In addition, more observations at higher airmass (airmass $> 1.5$) would allow us to more robustly test the airmass-dependency in the modified Young's equation (Eq.~\ref{eq:modyoung}). However, NGTS normally avoids higher airmass observations when attempting to acquire high-precision time-series photometry for astrophysical research.

In this work we have demonstrated that NGTS bright star observations reach the scintillation limit. We achieve this through a combination of improvements to the hardware and software relative to previous transit surveys. This is an example of the wealth of site characterisation data that lies hidden in archival datasets. This work can be extended to any observatory with telescopes that reach the scintillation limit and carry out high-precision time-series photometry. This provides an alternative method for characterising atmospheric conditions without any need for additional instrumentation or dedicated telescope time.

\section*{Acknowledgements} 

Based on data collected under the NGTS project at the ESO La Silla Paranal Observatory. The NGTS facility is operated by the consortium institutes with support from the UK Science and Technology Facilities Council (STFC) project ST/M001962/1.
The ECMWF data are generated using Copernicus Climate Change Service information [2021].
For accessing the MASS-DIMM data, this research has made use of the services of the ESO Science Archive Facility.
JO: We acknowledge the UK Research and Innovation Future Leaders Fellowship (MR/S035338/1).
JAGJ acknowledges support from grant HST-GO-15955.004-A from the Space Telescope Science Institute, which is operated by the Association of Universities for Research in Astronomy, Inc., under NASA contract NAS 5-26555.
JIV acknowledges support of CONICYT-PFCHA/Doctorado Nacional-21191829.

\section*{Data Availability}

NGTS FITS images are publicly available through the ESO Data Archive, access is described here: \url{http://ngtransits.org/data.html}
MASS-DIMM data are publicly available from the ESO Paranal Ambient Query Forms:
\url{http://archive.eso.org/cms/eso-data/ambient-conditions/paranal-ambient-query-forms.html}
The ECMWF dataset used in this work is publicly available at: \url{http://dx.doi.org/10.24381/cds.6860a573}

\bibliographystyle{mnras}
\bibliography{msc_paper}



\bsp	
\label{lastpage}
\end{document}